\documentclass[a4paper,twoside]{article}
\newcommand{\affil}[1]{$^{\rm #1}$}
\usepackage[authoryear]{natbib}
\bibpunct{(}{)}{;}{a}{}{,}
\usepackage{graphicx}
\date{} 

\title{\large\bf\flushleft Galaxy and Mass Assembly (GAMA): Optimal Tiling of Dense Surveys with a Multi-Object Spectrograph}
\author{\parbox{\textwidth}{\flushleft
\vspace{-0.5cm}
{\it Aaron Robotham\affil{A,P},
S.P.Driver\affil{A},
P.Norberg\affil{B},
I.K.Baldry\affil{C},
S.P.Bamford\affil{D},
A.M.Hopkins\affil{E},
J.Liske\affil{F},
J.Loveday\affil{G},
J.A.Peacock\affil{B},
E.Cameron\affil{H},
S.M.Croom\affil{I},
I.F.Doyle\affil{J},
C.S.Frenk\affil{K},
D.T.Hill\affil{A},
D.H.Jones\affil{E},
E.van Kampen\affil{F},
L.S.Kelvin\affil{A},
K.Kuijken\affil{L},
R.C.Nichol\affil{J},
H.R.Parkinson\affil{B},
C.C.Popescu\affil{M},
M.Prescott\affil{C},
R.G.Sharp\affil{E},
W.J.Sutherland\affil{N},
D.Thomas\affil{J},
R.J.Tuffs\affil{O}}\\
\vspace{0.4cm}
{\small \affil{A}\,(Scottish Universities Physics Alliance, SUPA) University of St Andrews, UK}\\
{\small \affil{B}\,(SUPA) University of Edinburgh, UK}\\
{\small \affil{C}\,Liverpool John Moores University, UK}\\
{\small \affil{D}\,University of Nottingham, UK}\\
{\small \affil{E}\,Anglo Australian Observatory, Australia}\\
{\small \affil{F}\,European Southern Observatory, Germany}\\
{\small \affil{G}\,University of Sussex, UK}\\
{\small \affil{H}\,Swiss Federal Institute of Technology (ETH), Switzerland}\\
{\small \affil{I}\,University of Sydney, Australia}\\
{\small \affil{J}\,University of Portsmouth, UK}\\
{\small \affil{K}\,Durham University, UK}\\
{\small \affil{L}\,University of Leiden, Netherlands}\\
{\small \affil{M}\,University of Central Lancashire, UK}\\
{\small \affil{N}\,Queen Mary University London, UK}\\
{\small \affil{O}\,Max-Plank Institute for Nuclear Physics (MPIK), Germany}\\
{\small \affil{P}\,Email: asgr@st-and.ac.uk}}}
\begin{document}
\twocolumn[
\begin{changemargin}{.8cm}{.5cm}
\begin{minipage}{.9\textwidth}
\vspace{-1cm}
\maketitle
\small{\bf Abstract:} A heuristic greedy algorithm is developed for efficiently tiling spatially dense redshift surveys. In its first application to the Galaxy and Mass Assembly (GAMA) redshift survey we find it rapidly improves the spatial uniformity of our data, and naturally corrects for any spatial bias introduced by the 2dF multi object spectrograph. We make conservative predictions for the final state of the GAMA redshift survey after our final allocation of time, and can be confident that even if worse than typical weather affects our observations, all of our main survey requirements will be met.


\medskip
\medskip
\end{minipage}
\end{changemargin}
]
\small

\section{Introduction}

Large redshift surveys are typically completed by observing with a Multi Object Spectrograph (MOS), obtaining spectra for many hundreds of sources simultaneously over large fields of view. The problem of how to optimise observing strategies to target sources distributed over some survey area with a given MOS, defining a field of view and number of simultaneous targets, falls into the ``area packing" class of problems. Much work outside of astronomy has been devoted to such problems \citep{megi84} which are usually intractable in a formal, provably-optimal, sense. In the case of the Anglo-Australian Telescope's (AAT) largest survey to date, the 2 degree Field Galaxy Redshift Survey \citep[2dFGRS,][]{coll01}, the survey was created in a manner that minimised field overlaps in order to maximise area (the target magnitude limit being $b_{j}=19.45$). This obviously had an impact on the target completeness, and the observations  had to be weighted in order to account for the local levels of incompleteness. At the other extreme is the 6 degree Field Galaxy Survey \citep[6dFGS,][]{jone04} that aimed for high levels of completeness within the local universe. In this case the filamentary structures (i.e.\ non-uniform overdensities) present on small scales necessitate extremely non-uniform tile coverage and potentially large amounts of overlap among tiles, target densities varying from 6 to 30 galaxies per deg$^{2}$. Hence the optimal strategy for tiling is closely linked to the scientific objectives of the survey, and a generic approach will not be appropriate for all requirements.

Fibre fed MOS instruments typically have a circular field of view (FOV), as seen for example in the 2 degree Field \citep[2dF,][]{lewi02}, 6 degree Field \citep[6dF,][]{jone04}, Sloan Digital Sky Survey (SDSS) Spectrograph \citep{york00}, Hectospec \citep{fabr05} and Hydra \citep{bard93}. Also typical is for survey regions to be rectangular in spherical coordinate geometry: recent examples include the 2dFGRS, Sloan Digital Sky Survey \citep[SDSS,][]{abaz09} and Millennium Galaxy Catalogue \citep[MGC,][]{lisk03,driv05}. This latter commonality is due to a number of allying factors: imaging CCDs used for input catalogues are almost always rectangular\footnote{The use of GALEX in WiggleZ \citep{glaz07} is a rare counter-example.} and survey boundaries and volumes are easier to consider when using spherical coordinate derived edges. Packing a shape best described in spherical coordinates into a Cartesian defined region is a non trivial task, and many approaches have been used in redshift surveys. Such packing problems are of wider mathematical interest because no provably optimal and rapid technique has yet been discovered \citep{megi84}. Instead every large survey tailors a tiling method in line with specific survey goals using a heuristic method. In this sense a heuristic method is one informed by knowledge of the problem at hand, the hope being the solution is not much worse than optimal. On top of the generic problem of efficient tile packing, spectroscopic surveys also have to contend with extremely non-uniform and complex selection functions within the tiles themselves. The major cause for the non-uniformity is object exclusion, either due to fibre collisions or slit overlaps.

In the case of 2dFGRS an approach close to hexagonal packing was used, where slight perturbations were made to a purely hexagonal grid of tile centres in order to better sample object densities. Since this survey was almost single pass (there was $\sim 30\%$ tile overlap), low completeness fields were not uncommon, an effect that was statistically adjusted for with observational weights. However, in the densest fields some targets will not have redshifts, and galaxy group assignments will not be as secure as in highly complete regions. The downside of such a regular approach is that all multi-fibre spectrographs will have structure or bias in their assignments and thus power will be added to (or removed from) certain frequencies in tangential modes of the power spectrum. The distribution of fibres is not only driven by the algorithm used to place them, but also the physical limitations of the instrument. Typically a fibre fed MOS is designed with fibres around the circumference in such a way that all fibres can reach the centre and few can reach locations at the edge, a scenario that makes radially-dependent targeting distributions inevitable. Even with the newest simulated annealing (SANN) algorithms available for AAOmega, radial assignment dependencies within each 2dF pointing exist \citep{misz06}. It is obviously important to try to compensate for such biases in any work that is concerned with clustering and structure, such as Galaxy And Mass Assembly \citep[GAMA,][]{driv09}, the latest large survey to use AAOmega on the AAT.

The spectroscopic element of SDSS \citep{blan03} used a heuristic algorithm that attempted to find an acceptable solution of a perturbed uniform grid of tiles, much like 2dFGRS. The algorithm aimed to utilise 90\% of the 600 available fibres on each tile, and similar to 2dFGRS the SDSS's median tile coverage for an object was 1 (both achieved a target density $\sim$100 galaxies per square degree). Minimum fibre spacings are $55''$ for the SDSS spectrograph, larger than the $40''$ distance for 2dF, thus an obvious limitation of SDSS is the full targeting and unbiased analysis of close pairs (a key science objective for GAMA, discussed in detail below).

Of recent surveys, the VIMOS VLT Deep Survey (VVDS from here) utilised the simplest approach to tiling \citep{bott05}. Effectively it placed tile centres on a fixed square grid with diagonal offsets used for the deeper component of the survey. Such an approach is possible when using VIMOS because of its mask-based grism spectrograph, giving it a square FOV better suited to tiling a square CCD photometric survey. The VVDS does not suffer from any radial selection bias, but due to the constraints imposed by slits cut into each mask it does possess complex selection effects such as the tendency to target a uniform spread of targets; highly clustered regions are hard to target since the slits necessarily avoid each other. Further complicating matters is a partially radial completeness bias, evident in the spectroscopic masks created for zCOSMOS \citep{knob09}. Whilst an interesting survey to note, such a survey design is not trivial to create with any of the fibre based multi object spectrographs discussed due to their circular FOV, and the complex radial bias this introduces.

Simulated annealing solutions of the tiling problem have been utilised in large area surveys with large amounts of structure present, most notably by the 6dFGS \citep{camp04}. Simulated annealing is a popular approach for many algorithmically insolvable problems and is, strictly speaking, a metaheuristic solution (i.e.\ choices have to be made about the element to be optimised and also the method of optimising). In simple terms the user must pick something to be minimised (or maximised), such as the total number of objects not assigned to a fibre after tiling the whole survey region. The user must also give the SANN algorithm variables to perturb (most obviously the right ascension and declination of the tile centres), and a rate at which it `cools' towards a solution. Typically these perturbations become smaller as the solution improves, and eventually an acceptable set of tile positions should be found. Packing problems lend themselves well to SANN since they can be tuned to find acceptable solutions rapidly, but they are non-deterministic algorithms (unlike the other heuristic approaches discussed) and are neither provably optimal nor stable (i.e.\ small variations to the problem to be optimised can produce radically different results). In the case of the 6dFGS, SANN is obviously much more effective than any sort of regular tiling because the projected target densities vary significantly and the survey area is large. The use of SANN reduces the number of sparsely populated fields and better samples overdensities where fields would be full.

Added to the complexities of these different approaches are the observational limitations for any survey as well as its scientific priorities. It will not be the case that all fields are equally observable in a large area survey (e.g.\ varying rising and setting time as a function of RA), but in a sufficiently small area survey it will often be the case that all parts of the survey field are effectively as observable as each other. Also, the end point of the survey will often be an unknown (i.e.\ weather dependent), so in many applications it is advantageous for the survey to be in a useable state as quickly as possible. With these extra considerations in mind, the philosophy that was applied to tiling GAMA was one where each tile would in some sense be the next most optimal tile, and every subsequent tile should make a significant impact towards achieving the GAMA survey requirements.

The GAMA redshift survey is one component of the multi-band GAMA survey project, and is the latest large survey to use the AAT's MOS facility. In this paper we explore the problems of tiling specifically for the GAMA survey, with the possibility of using the approaches discussed in future redshift surveys with characteristics in common with GAMA. In section 2 we outline the GAMA survey, and how the scientific goals for the project translate into survey requirements that our tiling algorithm must achieve. In section 3 we discuss in detail the different options to tiling that are appropriate for GAMA. In section 4 we apply the two most likely candidates for the tiling algorithm to the GAMA survey as it was left at the end of year 1, allowing quantitative judgements of the different approaches to be made. In section 5 we apply our chosen tiling algorithm to the data and present the state of the survey after year 2. Finally, conservative predictions are made for the state of the survey after year 3 observations based on tiling simulations.

\section{The GAMA Survey}

The GAMA project is a multi-band imaging and spectroscopic survey containing just under 144 square degrees of sky in three nearly identical $12^{\circ} \times 4^{\circ}$ areas centred on $9^{h}$ $+1^{\circ}$; $12^{h}$ $+0^{\circ}$ and $14^{h}30^{m}$ $+0^{\circ}$ (known as GAMA 09 or G09, GAMA 12 or G12 and GAMA 15 or G15). Future expansion to include two Southern $8^{\circ}\times6^{\circ}$ regions, to meet the survey requirements for measuring the halo mass function \citep{driv09}, is part of the design consideration. One of these southern regions may also be the focus of Australian Square Kilometre Array Pathfinder observations \citep{john08} in the proposed DINGO programme. Eventually all regions will be fully covered in FUV, NUV, $u$, $g$, $r$, $i$, $z$, $Y$, $J$, $H$, $K$ and far-IR, and will utilise imaging data from the SDSS, UKIRT Infrared Deep Sky Survey (UKIDSS), VLT Survey Telescope (VST), Visible and Infrared Survey Telescope for Astronomy (VISTA), GALEX and the Herschel Space Observatory. This imaging dataset is being complemented by a three-year redshift survey using the AAOmega spectrograph at the Anglo Australian Telescope (AAT). Observations allocated during 2008 (year 1) and 2009 (year 2) have been completed, with a third allocation of observing time during 2010 (year 3) remaining. The 2008 observations were made using a different approach to tiling (as discussed in detail below), and the tiling algorithms discussed here continue from the state it was left in then.

The GAMA survey is the latest in a long line of large galaxy surveys using the AAT to obtain redshifts (e.g.\ 2dFGRS and MGC), and is primarily designed to measure the halo mass function (HMF), with other scientific goals including an investigation of close pairs of galaxies (i.e.\ merging systems) and a fully dust corrected description of the galaxy luminosity function (LF) from the far-UV to the far-IR, along with the associated galaxy stellar mass function. The GAMA redshift survey aims to be exceptionally complete over the three large areas of sky described above. This requires careful planning in order to maximise the scientific output of the AAOmega instrument used to measure galaxy spectra \citep[see][and the AAOmega website\footnote{www.aao.gov.au/AAO/2df/aaomega/ aaomega\_manuals.html} for details]{shar06}.

In the case of the spectroscopic component of the GAMA survey, the requirement is for extremely high levels of completeness for all objects that are within our sample selection. This requires repeated observations for all areas of the survey, and thus tile placements become increasingly non-regular as the survey progresses in order to successfully target residual overdensities that appear.

The tiling algorithm used for GAMA must achieve a number of scientific goals (which have been translated into survey requirements) assuming conservative assumptions regarding observing time lost to weather. GAMA has strict primary targets, chosen so as to maximise our scientific return, and secondary goals to aim for upon completion of these.

\subsection{Survey Requirements}

Listed below are the primary survey requirements, which should be achieved by the end of the third year of observations at the AAT (assuming typical time loss due to bad weather and equipment failure). All references to completeness refer to the fraction of targets assigned at least one fibre compared to the number of objects in the input catalogue of targets. This does not mean all of these objects will eventually have redshifts (typically only 90--99\% of targets return a redshift), or that all of the targets are galaxies (e.g.\ our star/galaxy separation is not perfect, see Baldry et al. 2009 {\it submitted} for details).

\begin{itemize}

\item Flux limit: Fibre assignments for 99\% of targets with $r_{petro}\le 19.4$ in G09 and G15 and $r_{petro}\le 19.8$ in G12. Also $K_{AB}\le 17.5$ and $z_{model}\le 18.2$ in all three GAMA regions. For later reference, objects that satisfy at least one of these magnitude limits are main survey targets.

\end{itemize}

\noindent GAMA aims to be $99\%$ complete (or better) in terms of targeting for these three survey bands in each region. The $r$-band limits account for 114,780 of the 119,859 galaxies that meet these combined flux limits (95.7\%). Of the remainder 4,079 are provided by the $K$-band limit, with only 1,000 galaxies introduced to our sample by the addition of the $z$-band limit. The $r$-band limit was defined by our scientific goals for GAMA, and is a compromise between depth (deeper surveys have more objects per square degree), time available given the area GAMA is covering (only so many galaxies can be targeted) and the probable $S/N$ we can expect with AAOmega (redshift success rates drop off as a function of magnitude). The $K$-band limit was adopted to improve the quality of GSMFs obtained with GAMA, and was the deepest possible that kept the total number of required redshifts within achievable bounds. Finally the $z$-band limit was introduced because it is the reddest band available in SDSS and should ensure completeness in $r$ and $K$ for low surface brightness galaxies. For further details on the exact target selection used for GAMA refer to \citet{bald09}.

\begin{itemize}

\item Spatial completeness: 99\% of each region to be at least 80\%  targeting complete on the angular scale of $0.14^{\circ}$.

\end{itemize}

\noindent In order to improve the halo mass function to significantly lower masses than previously probed it is important that we have both high overall completeness (as defined above), as well as high levels of completeness on small spatial scales. Since the structures of interest are groups and clusters, the comoving physical scale of interest is $\sim 1$Mpc, and at $z=0.1$ (typical for high confidence systems) this subtends $\sim0.14^{\circ}$ (projected comoving distance when $H_{0}$=71 kms$^{-1}$Mpc$^{-1}$ assuming $\Omega_{m}=0.3$ and $\Omega_{\Lambda}$=0.7). For reliable estimates of velocity dispersions, and indeed for structures to be identified in the first place, a large fraction of potential members must have redshifts. In the case of very low mass groups (the type that we are most interested in) we require at least two redshifts to attempt a velocity dispersion (in the strictest sense this is true for the same reason we can measure the standard deviation of 2 data points, but more data are required to measure the velocity dispersion confidently). 80\% completeness means our expectation for a 3 object system is 2 or more redshifts, and 4 redshifts in a 5 object system. The desire that this level of spatial completeness is achieved in 99\% of each GAMA region is one of practicality, 100\% is obviously desirable, but 99\% is acceptable (i.e.\ we would not miss too many groups).

\begin{itemize}

\item Pair completeness: Fibre assignments for 99\% of galaxies within $40''$ of another galaxy.

\end{itemize}

\noindent Another scientific goal for GAMA is to thoroughly explore the merger rate of galaxies out to $z=0.5$. Since merging systems will necessarily be close on the sky, this obviously requires high levels of redshift completeness for galaxies with small angular separations. The value of $40''$ was chosen since this is the separation at which fibre collisions on 2dF become a significant issue. Measuring closely clustered objects on scales smaller than this limit is potentially difficult and must be approached as part of the primary survey observing strategy.

\subsection{Extended Survey Goals}

\begin{itemize}

\item Flux limit uniformity: Every 0.1 magnitude bin 99\% redshift complete for the magnitude limits given above.

\end{itemize}

\noindent Since redshift completeness is a function of flux (it is harder to obtain reliable redshifts for fainter objects) care should be taken so that our sample is not preferentially biased to brighter galaxies. This is a much harder target than achieving 99\% overall completeness, and since the effect can be corrected for later this is only considered to be a secondary survey goal. Should observing progress successfully, and assuming the requirements discussed above have been met, this could be an important survey goal in the latter stages.

\begin{itemize}

\item All galaxies should be observed with $-2^{h} \le$ HA $\le 2^{h}$ (where HA is the hour angle).

\end{itemize}

\noindent Whilst it is desirable that every galaxy is observed at zenith for the entirety of the integration period, this is obviously not possible. A sensible constraint for GAMA is that all objects should be observed within $2^{h}$ of the meridian in order to keep the air-mass down, but in exceptional conditions this constraint may have to be omitted for reasons of practicality. It is generally true to say that when one of G09, G12 or G15 is observable all galaxy positions are equally acceptable within a region, the exceptions being at the extreme of our allowed HA range.

\begin{itemize}

\item Reobservation of all targets for which we failed to obtain a redshift.

\end{itemize}

\noindent A large fraction of redshift failures will be caused by effects unrelated to the true viability of a target. For instance partial cloud cover during observation or fibre positioning errors both conspire to reduce the amount of flux entering a target fibre, and since the chance of obtaining a redshift is proportional to the $S/N$ this will mean fainter objects are more likely to be classed as a failed target. So as not to introduce any unwanted targeting bias to the GAMA survey, we ideally should observe all failed targets at least twice. As well as giving the object a chance to be observed in a more favourable plate position and better weather conditions, we can use the summed integration time even if $S/N$ is low in the reobservation. Thus our redshift survey should be minimally biased by flux.

\subsection{GAMA Survey to Date}

Beyond achieving the requirements and goals stated above, a complicating factor for the tiling algorithm to be used is that it must continue the GAMA survey from how it was left at the end of the first year of observations. Due to tight time constraints, the year 1 tiling of GAMA was implemented using a simplistic gridding system where each region was divided into three rows and eight columns, with the divisions being lines of longitude and latitude in spherical coordinates. Each vertical box-edge was adjusted in right ascension (RA) until all boxes in a row contained a similar number of targets, then objects were extracted into 2 separate catalogs containing half the targets each. The aim of the first year was to try to observe each box twice, a feat that was nearly achieved due to three extremely successful weeks of observations at the AAT. Whilst this returned a fantastic number of redshifts ($\sim 51,000$) it became apparent that the distribution of objects with redshifts betrayed clear signs of their gridded origin; an effect of the configuration routine for the 2dF. This routine, known as {\sc configure}, is supplied to observers at the AAT in order to convert lists of desired targets into valid fibre locations on the 2dF, and is the closest interface observers have to the eventual distribution of fibres \citep{misz06}.

\begin{figure}[h]
\begin{center}
\includegraphics[width=8cm, angle=0]{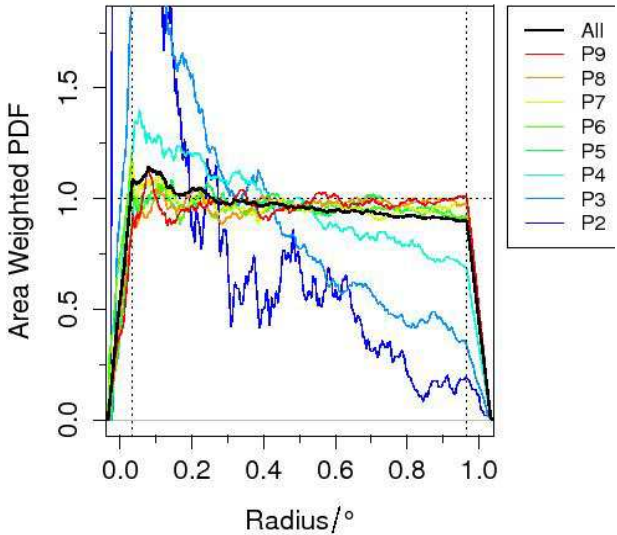}
\caption{\footnotesize The radial bias of the simulated annealing algorithm used in the AAT 2dF {\sc configure} software. 300 simulations of a random uniform 2dF region were configured with the tiling software, where 600 targets were randomly assigned a priority level between 1 and 9 (higher number indicates higher priority), and 378 fibres were working. All densities are weighted by area, thus no radial gradient would be a uniform distribution in this plot. All combined priority levels are plotted (black line) as well as all priority levels from 2--9 (blue--red). A rectangular density kernel was used with a bandwidth of 0.02. Horizontal dotted line denotes the uniform distribution. Vertical dotted lines denote regions beyond which edge effects render the densities meaningless because the bandwidth samples outside of the physical limits of the 2dF.}\label{priorplot}
\end{center}
\end{figure}

Whilst the newest versions of {\sc configure} (a GAMA-specific version 7.10+ was used throughout) offers vast improvements over older routines, and produces much less pronounced spatial features, it still possesses a clear radial gradient. Evidence of this gradient can be found in Figure \ref{priorplot}. This plot shows the probability of targets obtaining a fibre as a function of distance from the 2dF centre. As well as demonstrating the general tendency for a random set of targets to have a central bias (the black line in the plot), different {\sc configure} priority levels were investigated separately, where a higher number (maximum of 9, minimum of 1) indicates the simulated annealing algorithm tries harder to put a fibre on a target. Radial effects are not evident, or are very small, for high priority levels, but it is clear from Figure \ref{priorplot} that the radial distortion becomes extremely noticeable for low priority objects for the simulations conducted here.

The effect of the algorithm is to return a more uniform distribution for the highest priority targets, at the expense of lower priorities. The result of this fibre assignment gradient is that given a region that has an even distribution of targets within the FOV, objects in the centre, especially those assigned a low priority, are more likely to be allocated a fibre than similar priority targets near the edge. This is an almost unavoidable effect since many more fibres are able to reach central targets. At the extreme, an object exactly in the centre of a field is reachable by all 392 fibres (400 minus the 8 guide fibre bundles), but one at the extreme edge of the field (directly in front of a fibre) might only by reachable by 1. 

In the example presented here, all priority 5 targets and higher could have been assigned a fibre in theory. This means that purely by virtue of assigning fibres to a large fraction of these targets a close to uniform distribution is assured, and hence the gradient is much more evident for priority level of 4 and below. As a guide to the gradient expected if all targets possess the same priority, the combined distribution is the most indicative (black line). Thus assigning all targets to a high priority will not eliminate the gradient, but the most undesirable features will always affect the lowest priority targets more.

\begin{figure}[t]
\begin{center}
\includegraphics[width=8cm]{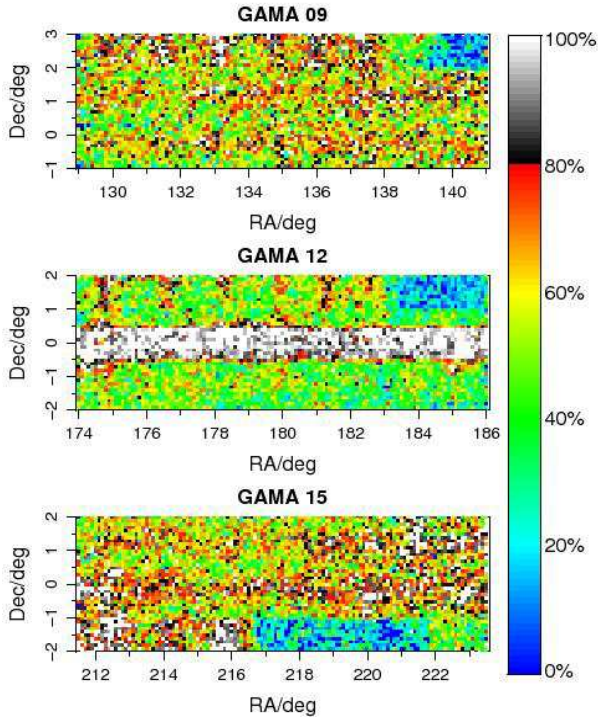}
\caption{\footnotesize The state of the GAMA regions after the first year of data. The plots describe survey redshift completeness inside a circular top-hat kernel with a diameter of $0.14^{\circ}$ This was chosen since it is the angular extent of a 1 Mpc system at $z\sim0.1$ assuming a $\Lambda$CDM cosmology and $H_{0}$=71 kms$^{-1}$Mpc$^{-1}$, and thus represents the group/cluster scale. Blue through to red represents 0\%--80\% completeness, whilst black through to white represents 80\%--100\% completeness. One of the main survey goals is that 99\% of the pixels in each GAMA survey area are 80\% complete, i.e.\ this plot is 99\% grey-scale.}\label{gamastart}
\end{center}
\end{figure}

The impact of such a radial selection function on data which is gridded in a Cartesian manner should be clear: corners are under-sampled compared to all other regions. This effect was exacerbated in the first year GAMA data because gridded subsets were observed twice. Figure \ref{gamastart} is a plot of local completeness, showing the fraction of main survey targets observed inside a circular top-hat of radius $0.14^{\circ}$ (the local completeness scale stated in the survey requirements). The central light strip in G12 is due to a deeper survey limit for this region, $r_{petro}\le19.8$ here compared to $r_{petro}\le19.0$ or $r_{petro}\le19.4$ for all other targets in year 1 (the use of these limits is discussed in detail below). Ignoring this strip, the next obvious feature is periodicity in completeness, demonstrating the clear Cartesian residual embedded in the data after the year 1 strategy. This coherent regular structure is due to radial effects in the {\sc configure} software. The most obvious features are long, highly complete regions that are at the same declination in GAMA 09 and GAMA 15 (the central strip in GAMA 12 is by design, as discussed above). 

Running orthogonally to these strips in right ascension are periodic strips in declination. Due to the target boxes being shuffled in right ascension, these strips do not necessarily span the full range of declination, but they are particularly obvious at the top of G09 and G12, and the bottom of G15. The extremely blue (incomplete) regions are those not visited during GAMA year 1---the reason these regions are not perfectly blue is because various older surveys (e.g.\ 2dFGRS and SDSS) already provide redshifts for a small fraction of GAMA targets here.

As well as needing to consider issues regarding the removal of non-cosmic structure from our completeness map, the year 1 GAMA survey was conducted with different magnitude limits to those now required. These were used in order to increase the scientific return from the first year of spectroscopic data, and should not negatively impact the survey from this point. The major difference from the GAMA survey requirement magnitude limits stated above is that only an $r$-band petrosian magnitude was used, and the limit was $r_{petro}\le 19.0$ in G09 and G15, and a mixture of $r_{petro}\le 19.0$ and $r_{petro}\le 19.4$ in G12 (due to the excellent weather G12 was extended in overall depth midway through year 1), with the addition of the deeper strip in G12 limited to $r_{petro}\le19.8$ (this strip is obvious in Figure \ref{gamastart}). Since there are two more years of observations to be made this selection effect should not be difficult to compensate for in the long term, and part of the reason our first extended survey goal is to achieve equal redshift completeness as a function of magnitude.

\section{Tiling Options Explored}

In algorithmic terms the approach desired for tiling GAMA from its post year 1 state is a type of heuristic greedy algorithm \citep{corm90}, where the tile about to be put down maximises some property of the survey, and in the longer term the task of tiling is not made too much harder by this greediness. Such an approach is both desirable and possible due to the extremely high object density required for the GAMA survey. This means the problem is contrary to the type applicable to low spatial density redshift surveys (e.g.\ the 6dF survey) because on average the number of 2dF tiles placed on a given area will be extremely high (conservative estimates suggest every position will be contained within at least six separate tiles), rather than deliberately low (i.e.\ minimally packed).

On a slightly separate issue, because the GAMA survey is particularly interested in low mass halos it is absolutely essential that highly clustered objects are attacked in an aggressive manner. There are constraints on this process however since the 2dF has physical limitations on how close together fibres can be placed. This problem can be solved by repeatedly observing clustered regions, and making sure the ``worst offending'' objects in clustered regions are observed as early as possible in order to achieve the tiling requirements.

The issue of tiling is interwoven with the problem of assigning fibres to targets on the 2dF. The program used for assigning fibres on the 2dF instrument ({\sc configure}) has been continually upgraded since its introduction 10 years ago, and now the algorithm of choice is based on simulated annealing of the fibre allocations. Whilst this approach offers massive advantages over the older Oxford and Taylor algorithms \citep[see][for details]{misz06}, it is non-deterministic. Every time the configuration is attempted a different solution will almost certainly be found (a feature to be added to {\sc configure} is the option of setting the random seed, but a small perturbation in the input file will still create a radically different solution). Since a typical configuration time is of the order 10 minutes it is computationally challenging to incorporate the fibre assignments into a long term optimisation approach to tiling, be this SANN or quasi-Newton BFGS (Broyden, Fletcher, Goldfarb and Shanno) optimisation of the tile positions \citep[for a discussion of multidimensional optimisation algorithms see][]{noce06}. For sparse surveys such as 6dF, where there is little tile overlap for the most part, this will not present such a problem since a given object is typically only in one tile, but for GAMA it rapidly impacts on the efficiency of the tiling. The other issue with optimising for all of the tile positions in a survey such as GAMA is that it offers no insight into where it would be best to place the next tile since this would mean optimising for $N_{tile}!$, where $N_{tile}$ is the number of tiles (i.e.\ every possible tile ordering). For 50 tiles this would mean $\sim10^{64}$ full survey configurations, and this is assuming the tile positions are already optimal. This makes the problem highly intractable computationally, instead the standard approach (e.g.\ 6dF) is to make all potential tiles an equally good option.

When the survey area is extremely large (i.e.\ only a small fraction of it is observable at a given moment) producing a large number of equally good target fields makes a lot of sense since it is hard to predict exactly which region will be within the required zenith distance range when observations start, the typical advised limit on a given field being $\pm2^{h}$ (hence this being a survey goal). In the case of GAMA, a 2dF tile can be placed in any part of the survey sub-regions, so we are free to place the next tile in the most optimal position. Since the longest a GAMA region can be observed for whilst remaining inside the hour-angle limits is $4^{h}48^{m}$ (the extra $48^{m}$ comes from the RA length of each GAMA region), the next GAMA region will always be at a smaller (more desirable) hour-angle before we are limited by RA within the current region. The exceptions to this are the first and last fields of the night, where it might be necessary to limit our observations to the survey region extremes in order to maximise observation time.

Bearing in mind these competing factors the final matter that must be decided is what aspect of the survey should be improved with each tile used. The two most obvious possibilities, based on the survey requirements discussed in the previous section, were the number of redshifts obtained (hereafter referred to as {\it greedy}), and the spatial completeness of the survey (hereafter referred to as {\it dengreedy}). The former case would simply involve determining which region of the survey has the greatest number of high priority targets within a two degree FOV, regardless of any other information. This would be the crudest type of greedy algorithm, in the mathematical sense, because all each tile cares about is where the densest collection of targets is. The reason this could become too crude is because even at the mid stage of the survey there will be multiple places in the survey region that contain far more targets within a 2dF tile than there are fibres, and whilst each of these tile locations would improve the total completeness of the survey by the same degree, they will not necessarily improve the spatial completeness by the same amount. The {\it {\it greedy}} algorithm might accidentally pick the location that improves the spatial completeness the most, but only a small fraction of the time. Hence always placing the tile centres at the densest point might be too greedy given our survey requirements.

The {\it dengreedy} approach of improving the spatial completeness is slightly more subtle. It works by choosing tile centres based on which location in the survey (when sampled with the 2dF) is the least spatially complete, regardless of how many targets are available. Whilst sounding potentially disastrous, allowing the tile centres freedom regardless of the number of targets works very effectively. Given the 2 degree FOV of the 2dF, the large scale structure of the universe introduces relatively small variations in the homogeneity of our target galaxies. By design, spatial optimisation achieves angular completeness faster than the purely greedy approach, but it does typically return fewer redshifts after a given number of tiles. Since the main scientific goal of GAMA is to measure the halo mass function for very low mass systems, which requires high spatial completeness, this is not necessarily a terrible compromise. It should be noted that {\it dengreedy} still generally favours regions missing the most redshifts (given the local variability of the large scale structure), but since the algorithm works specifically to level the spatial completeness it will often find quite different tile position solutions given the same survey state.

\begin{figure}[h]
\begin{center}
\includegraphics[width=8cm, angle=0]{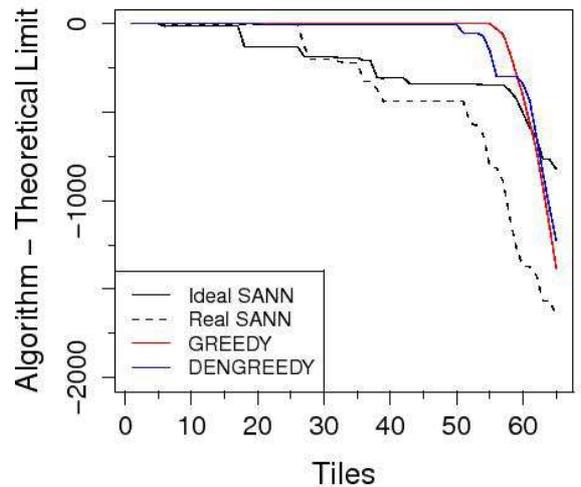}
\caption{\footnotesize Comparison of different tiling approaches. Simulated annealing (SANN), {\it greedy} and {\it dengreedy} approaches to tiling are simulated on identical data. 22,000 objects are randomly distributed inside a $12^{\circ}\times4^{\circ}$ area and 350 objects (at most) are removed each time a simulated observation is completed. The plot shows the cumulative difference in objects extracted from the maximum possible as a function of tile number. If all possible tiles are observed the optimal type of tiling is a variety of simulated annealing (Ideal SANN), but this is only more efficient when removed targets are predictable and nearly every expected tile is used (Real SANN performs significantly worse). For list of priorities used in the GAMA survey, see Table \ref{priortab}}\label{compmethod}
\end{center}
\end{figure}

A simplistic comparison of {\it greedy}, {\it dengreedy} and SANN (the most implementable type of full tile position optimisation when the number of free parameters is large since it is resistant to local minima) is made in Figure \ref{compmethod}. This shows the cumulative difference between the total number of targets acquired and the maximum possible as a function of tile number. 22,000 objects were randomly generated in an area the same size as a GAMA region; this number chosen because it is roughly the number of objects left to target in G12. Since the tiling imprint dominates the target object structure rapidly, a uniform distribution of targets is adequate for comparative purposes. The actual {\sc configure} program is not used (this would be too time consuming), instead 350 objects are randomly removed from a 2 degree FOV (without replacement) from the survey area for each tile, and the plot shows the cumulative difference in objects targeted as a function of tile number. The ideal simulated annealing (Ideal SANN) removes objects from each tile in a consistent manner (to simulate the output of {\it configure} being predictable), and also has a specified number of tiles to use (65). {\it greedy} and {\it dengreedy} on the other hand attempt to improve the total survey completeness or spatial completeness as much as possible with each subsequent tile.

To reflect how the random distribution of targets produced by {\sc configure} can affect the efficiency of simulated annealing, a variation of this tiling was made (Real SANN) which uses the same tile positions as Ideal SANN, but randomly selects 350 objects. This removal of objects is also done without replacement, but it is non-deterministic and thus will not return the same object assignment solution as Ideal SANN. Clearly this has a significant impact on the efficiency of the tiling, and it means simulated annealing goes from being the most effective approach (when targeted objects are deterministic and nearly every tile generated is used) to the worst. Interestingly, {\it dengreedy} achieves higher levels of completeness than {\it greedy} towards the end of these simulations. These are simple comparisons, but they do highlight the issue of how simulated annealing will find good solutions only when the inputs to the problem are precisely known. If a survey finishes a few tiles sooner than expected due to bad weather (a realistic prospect for many surveys) then the gains brought by SANN are lost, and equally if there is a non-deterministic black-box contained within the problem to be optimised (in this case {\sc configure}) then the solution could be far from optimal.

Since the number of tiles remaining for the GAMA spectroscopic survey is unknown, and the small survey area lends itself well to observing the next best position at nearly all times, a decision was made at an early stage to concentrate efforts on investigating the {\it greedy} and {\it dengreedy} algorithms. This means approaches that attempt to optimise for all tile positions (in this case SANN, but includes any type of multidimensional optimisation routines such as BFGS) will not be discussed further since they cannot truly optimise for a non-deterministic configuration routine and an unknown number of remaining tiles. The other weakness of total survey optimisation is that it cannot properly compensate for the subtle effects of fibre targeting gradients discussed in the previous section, whilst a tile-by-tile type of optimisation will continually make small adjustments based on exactly these effects.

\subsection{Which Type of Greedy?}

\begin{figure}[t]
\begin{center}
\includegraphics[width=8cm]{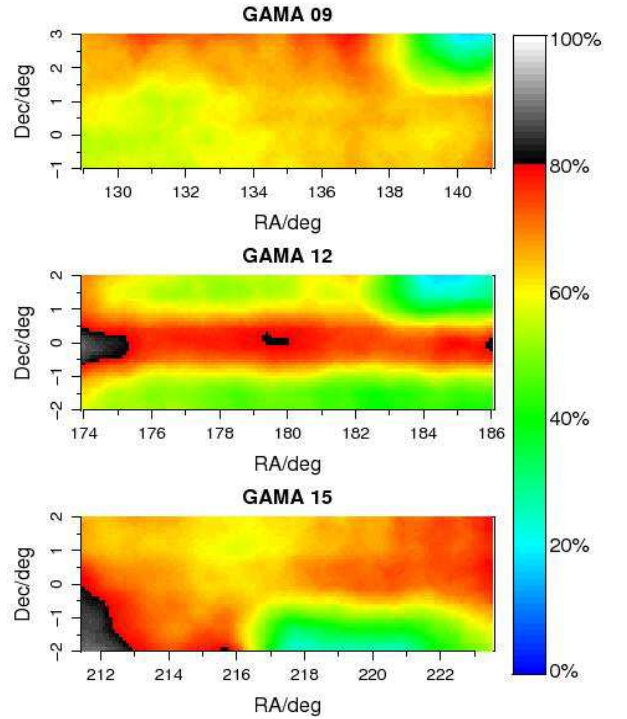}
\caption{\footnotesize The spatial completeness of the GAMA regions after the first year of data. The plots describe survey completeness inside a circular top-hat kernel with a diameter of $2^{\circ}$. See Figure \ref{gamastart} for further details of the completeness metric.}\label{gamastart2dF}
\end{center}
\end{figure}

To determine the optimal position of the next tile both the {\it greedy} and {\it dengreedy} approaches were investigated thoroughly. The {\it greedy} algorithm will simply choose the tile location that has the most main survey targets within it, for instance in G09 this would be in the centre of the unobserved region in the top-right (see Figure \ref{gamastart}). The {\it dengreedy} algorithm, however, would not pick exactly the same location. Because it convolves the targets with the full two degree FOV, the least complete point in the survey tends to be at the extreme edge of the survey region when there is a large incomplete area. This is clear in Figure \ref{gamastart2dF} where the GAMA incompleteness maps have been convolved with the full 2 degree FOV. The most incomplete point in G09 is the extreme top-right corner when considered in this manner.

It was realised that allowing the field centres to move to such extremes would, in the long term, be detrimental to the survey. The most serious concern is that too few objects might be selected to use $\sim100\%$ of the available fibres, and even if there were plenty of targets in the field, the Cartesian geometry could reduce the fraction of targets successfully assigned. The obvious solution is to put mild limits on how close to the survey edge the tile centres are allowed to be, effectively limiting the part of the survey that can provide a minimum in the completeness map. Simulations were conducted on G09 to ascertain the ideal distance to use, the results suggesting that any buffer between $0.3^{\circ}$--$0.5^{\circ}$ improves the tiling efficiency (survey requirements met faster), and $0.4^{\circ}$ appeared to be about optimal (survey requirements obtained 2 tiles faster than without a buffer). These buffer zones should not be enforced when the number of targets remaining is very small (hundreds within a GAMA region) because the extreme region edges will often be the best place to place a tile.

\begin{figure}[t]
\begin{center}
\includegraphics[width=8cm, angle=0]{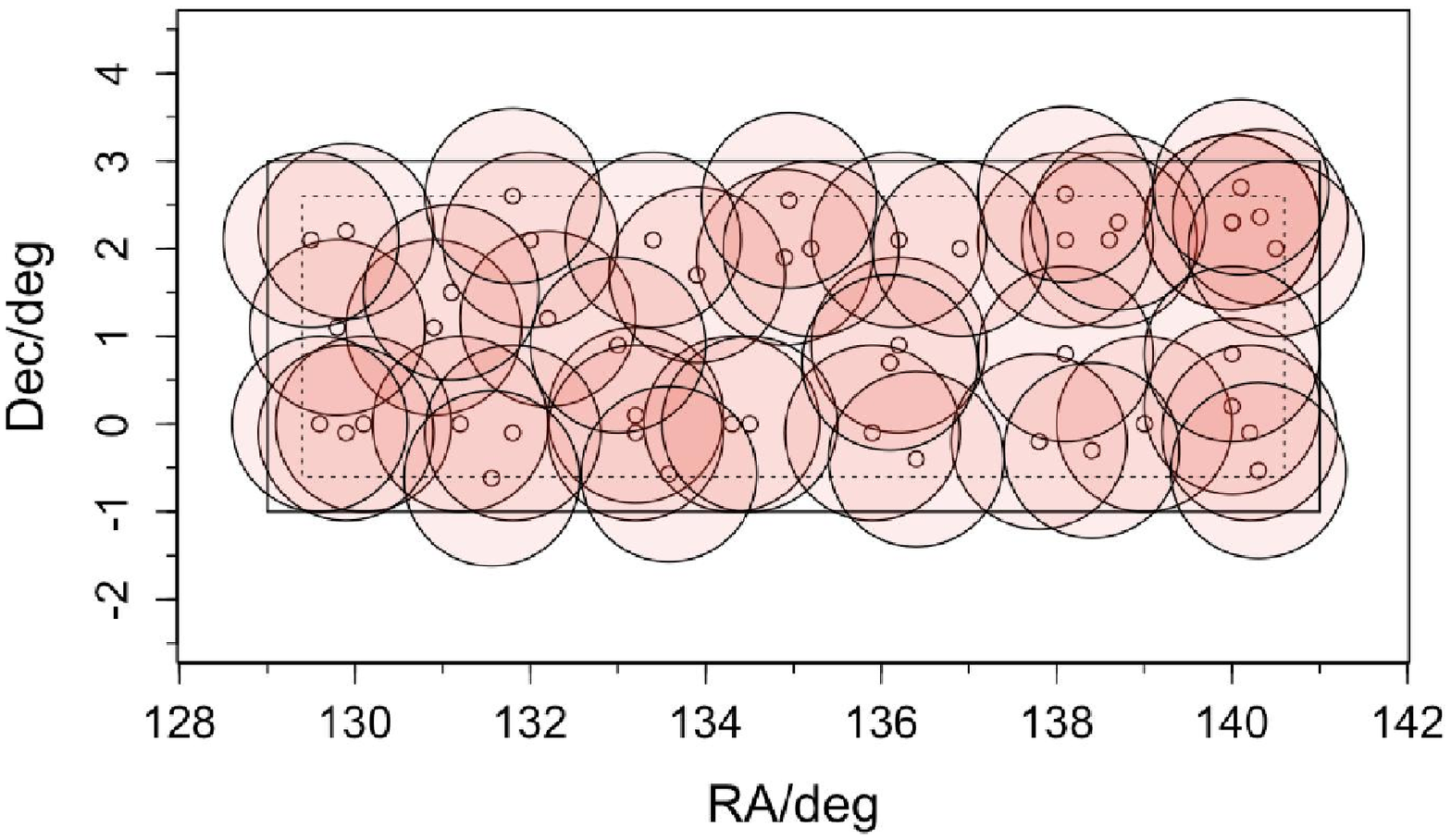}
\includegraphics[width=8cm, angle=0]{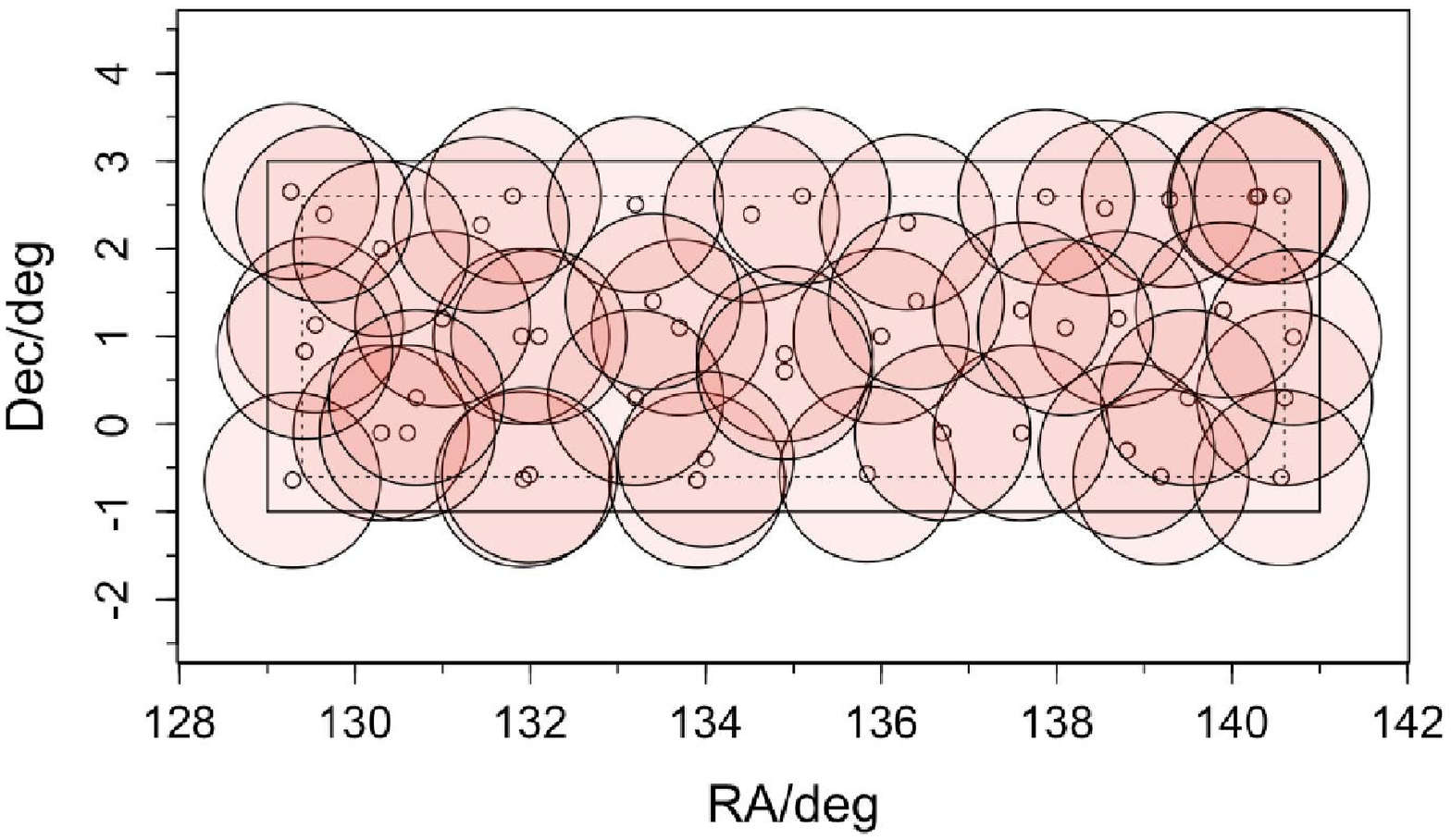}
\caption{\footnotesize Plots demonstrating the differing distribution of 2dF tiles when GAMA 09 has achieved 99\% completeness, using both the {\it greedy} (top) and {\it dengreedy} (bottom) approaches for the tiling metric. The dotted line, in both, plots indicates a $0.4^{\circ}$ tile centre buffer. It is clear that the {\it greedy} algorithm typically positions tiles a large distance from the survey edge, whilst {\it dengreedy} often places tiles right up to the survey buffer limit. Both approaches concentrate tiles on the least complete regions of GAMA 09 (as seen in Figure \ref{gamastart}), hence the large number in the top-right region of GAMA 09. {\it dengreedy} produces better packing, which translates to less overlap between tiles.}\label{gamatiles}
\end{center}
\end{figure}

Such a positional limitation is not necessary for the {\it greedy} algorithm because it will rarely be the case that more targets will be contained within a FOV at a region edge than slightly inset. Generally a {\it greedy} tile centre will be nearly $1^{\circ}$ from a survey edge to maximise the number of targets within. These subtle effects can be seen in the plots of Figure \ref{gamatiles}, which show the positions of tile centres using both the {\it greedy} and the {\it dengreedy} approaches for the tiling metric using survey buffers (the tile centres inside the buffer zone are due to the caveats discussed above). The {\it greedy} algorithm generally positions tiles much further inside the survey limits, the average distance of each tile from the centre of GAMA 09 is $3.54^{\circ}$ for {\it greedy} and $3.66^{\circ}$ for {\it dengreedy}. The consequence is that there is more overlap between tiles using {\it greedy}, and that it takes longer for every part of GAMA 09 to have been contained within a two degree FOV once. Both plots show the positions of the tiles that bring the survey completeness up to 99\%, which in these simulations happens to occur after 48 tiles for both {\it greedy} and {\it dengreedy} (run to run, the exact number of tiles will differ due to the random nature of the simulated annealing used in {\sc configure}).

\begin{figure}[t]
\begin{center}
\includegraphics[width=8cm, angle=0]{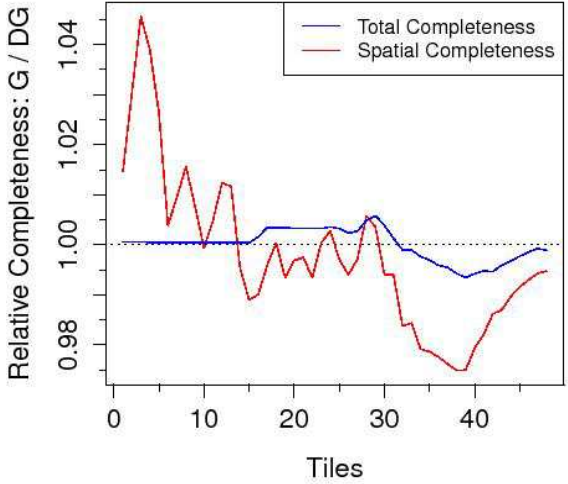}
\caption{\footnotesize Plot comparing {\it greedy} and {\it dengreedy}. The y-axis shows the relative total and spatial completeness: the {\it greedy} completeness divided by the {\it dengreedy} completeness. When this ratio is greater than 1 the {\it greedy} algorithm is doing a superior job, and the reverse is true when the ratio is below 1. The data is plotted up to tile 48 (when both algorithms achieve the required spatial completeness). Whilst much of the early tiling favours the {\it greedy} algorithm, {\it dengreedy} is clearly doing a better job of improving spatial and total completeness when we are within ~15 tiles of the survey's end.}
\label{relcomp}
\end{center}
\end{figure}

The major advantage of using {\it dengreedy} over {\it greedy} is in the latter stages of the survey when approaching high total completeness (remembering our requirement is 99\%). As a qualitative example, whilst the {\it greedy} algorithm is naturally biased towards large clusters that are missing, say, 20\% of potential members, {\it dengreedy} will be drawn towards less dense regions containing numerically poor groups missing, say, 25\% of potential members. Whilst the large cluster may be missing more objects in total, its dynamics will already be reliably measurable at 80\% completeness. The more tenuous small groups require very high levels of completeness to confidently apply grouping algorithms (e.g.\ Friends-of-Friends), and in order to construct the halo mass function down to exciting new levels it is these systems that are the key. As should be expected, {\it dengreedy} achieves our spatial completeness targets (99\% of the survey area is locally at least 80\% complete) faster than {\it greedy} (46 tiles, compared to 48, in these simulations). Figure \ref{relcomp} demonstrates the the long term superiority of the {\it dengreedy} algorithm clearly. When we are close to the end of the survey (within ~15 tiles) {\it dengreedy} returns consistently better total and spatial completeness. This means should our survey be extremely hindered by bad weather or technical problems, the data set will be much more complete. Based on this reasoning, the tiling algorithm that we selected for continuing the GAMA spectroscopic survey was {\it dengreedy}.

\section{Tiling Algorithm Implementation}

Having chosen {\it dengreedy} as our tiling method, we must now consider a number of issues that can significantly impact the efficiency of our survey regardless of the tiling algorithm to be used.

\subsection{Priority Bumping}
Since one of our survey requirements is high completeness for close pair targets, an issue that had to be addressed was fibre collisions hindering the rate at which clusters can be maximally sampled. Typically two fibre buttons can be no closer together than $40''$ (the actual exclusion geometry is more complex, but this is deemed an appropriate estimate on the AAOmega website), which at $z=0.2$ (approximately the median redshift of GAMA) corresponds to 131 kpc. A compact group might have numerous galaxies closer together than this distance, even ignoring projections that render any system more closely packed when observed. Added to this, one of the primary scientific goals of GAMA is an analysis of merging galaxies and close pairs, so placing fibres on a large fraction of such pairs is vital. The only way to overcome the problem of fibre collisions is by re-observation of the same region of sky, a certainty in the GAMA survey. Thus in order to observe clustered regions as efficiently as possible, an aggressive approach to close pair targeting was used.

For each tile generated a collision matrix of all the main survey targets is created. From this the worst offending target (i.e.\ target that is within $40''$ of the most other targets) is found, and its priority level is increased by 1. This makes it much more likely the {\sc configure} program will place a fibre on it in the tile being created. Furthermore, in order to improve the chances these colliding targets are successfully assigned a fibre, all the objects that they are interfering with are removed from the list of potential targets for the tile being made. This last step is important since all the highest priority targets would otherwise be in regions that are difficult to configure, and the simulated annealing algorithm will often cool to a solution before a large fraction of these targets are assigned a fibre. With the worst offending target bumped up one level of priority and the interfering targets removed, the next worst offending collider is found and the process repeated until no objects closer than $40''$ remain in the sample of interest.

By following this process for every tile made, usually 100\% of the highly colliding targets are removed  each time, and consequently as the survey approaches high levels of completeness we are not left with pockets of targets that require multiple configurations. The effectiveness of this aggressive approach to targeting clusters is clear from simulations conducted for the GAMA 09 region: using priority bumping means 99\% spatial completeness is achieved with 46 tiles (from the survey state at the end of year 1 using {\it dengreedy}), however if no priority bumping is used this same level of completeness typically requires 2--3 more tiles. Obviously the local spatial completeness considers angular regions much larger than the $40''$ collisions being targeted by the priority bumping, but the long term rewards of the approach seem clear.

\begin{figure}[h]
\begin{center}
\includegraphics[width=8cm, angle=0]{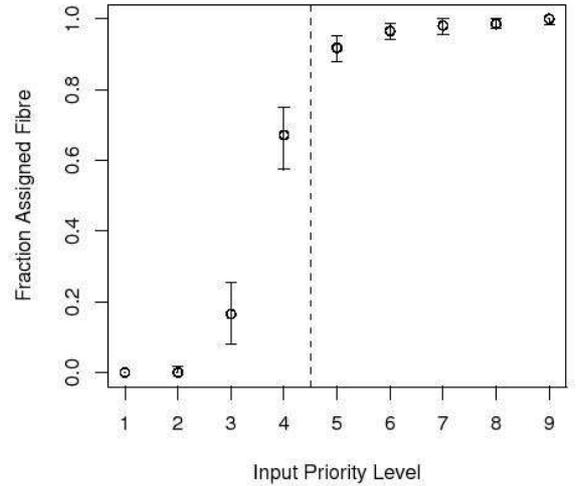}
\caption{\footnotesize The fraction of fibre assignments to potential targets for different priority levels. 300 Monte-Carlo simulations were made, where 600 objects were uniformly distributed in spherical coordinates within a 2 degree FOV and assigned a priority level between 1 and 9. There were 378 fibres available for each configuration. The dotted line indicates that priorities 5 and above always within the highest priority 378 objects, so in theory these higher priorities could all be complete. The error bars indicate the 15.9\% and 84.1\% quartiles for the assignment fractions from the 300 simulations, so reflect $1\sigma$ errors.}\label{priorbar}
\end{center}
\end{figure}
\subsection{Priority Levels}

When constructing input files for {\sc configure} (files with the {\bf .fld} extension) care must also be taken with how priorities are assigned to targets. Figure \ref{priorplot} demonstrated how the highest priority targets are also those with the least radial bias, whilst Figure \ref{priorbar} shows that even when there are plenty of fibres available, higher priority targets will obtain better completeness. The main GAMA survey was awarded priority levels of 6, 7 and 8. Priority 6 objects are main survey objects that have been observed once but a redshift was not obtained. Since weather conditions and the location of the target on the 2dF drive redshift success rates (fibre placement errors occur as a function of tile position, hence $S/N$ and redshift success), it is prudent to observe such failures more than once, and these come back into the target list at a lower priority than the unobserved objects. Priority 7 objects are main survey targets that have not been observed and that are not highly clustered, or priority 6 objects that are highly clustered and have had their priority bumped. Priority 8 is reserved for highly clustered priority 7 targets that have had their priority level bumped up. Priority 9 is reserved for spectral standards (only 3 per field) and emergency additions--- although this back-up functionality was not required. To guarantee that every fibre is used (and to make headway on any deeper redshift survey in the same region), filler targets were created and these targets were assigned lower priorities. The full list of priority levels and object types for year 2 onwards can be found in Table \ref{priortab}.

The priorities assigned to targets were different between the two years. In year 1, the targets consisted only of the $r$-band selection with $r_{psf}-r_{model}>0.25$ (there was insufficient UKIDSS coverage at the time), without an already known redshift. The priorities were from high-to-low: (i) $r<19.0$; (ii) $19.0<r<19.8$ in G12 within $\pm0.5^{\circ}$ of the celestial equator (creating the central strip clear in Figure \ref{gamastart}); (iii) $19.0<r<19.4$ in G09 and G15, and remaining $19.0<r<19.8$ in G12.  In addition, clustered targets in any of these categories were given a higher priority.

\begin{table}[h]
\begin{center}
\begin{tabular}{l|l}
Priority & Object Type\\
\hline\hline 9 & Spectral Standards \\
\hline 8 & Clustered P7 \\
\hline 7 & Main Survey/ Clustered P6 \\
\hline 6 & Failed Main Survey \\
\hline 2--5 & Filler Targets \\
\hline
\end{tabular}
\caption{\footnotesize Priority table. Main survey targets are within the main GAMA regions and have $r\le 19.4$ for G09/G15, $r\le 19.8$ for G12, or $K_{AB}\le 17.5$ or $z_{model}\le 18.2$ for any region. The priority 2--5 filler targets use $r\le 19.8$ for G09 and G15, and has any one of $g_{model}\le 20.6$, $r_{model}\le 19.8$ or $i_{model}\le 19.4$. Selected fillers also cover an extended survey area using the main survey magnitude limits, the GAMA regions beoming $14^{\circ}\times4.5^{\circ}$ strips. Also used as filler objects are objects that either have poor quality AAOmega spectra, or are missing it altogether because the redshift comes from an older survey.}\label{priortab}
\end{center}
\end{table}

To create a configuration input file 600 targets are drawn from the input catalogue. This number was chosen since it allows enough overhead for every fibre to be used, whilst remaining small enough to keep configuration times down. To achieve this number of targets in the {\bf .fld} file all the priority 8 targets within the FOV are extracted, if there are more than 600 then a random sample of 600 is taken, if there are less than 600 then all of them are put into the {\bf .fld} file. Assuming, for example, there are 150 priority 8 objects, then 450 spaces remain in the file. Next all the available priority 7 objects are extracted, again if there are fewer than 450 all of them are used, or else a random sample of 450 is taken and used. This process is repeated down to the priority level that can fill all the remaining slots, or until all targets within the FOV have been used. In practice the former condition is always reached first.

After extensive testing it was decided that priority 6,7 and 8 targets would be used to determine the locations of tiles, where all three priority levels carry the same weighting when calculating the completeness within a 2 degree FOV for {\it dengreedy}. This means objects that have spectra, but were not of high enough $S/N$ to obtain a reliable redshift, are allowed to influence the positions of the tiles. This seems reasonable when considering that the redshift success rate within a field can reach 100\% when the seeing and weather are ideal, but drop considerably when conditions worsen, so in order to not introduce a temporal bias these redshift failures should be re-observed and be allowed to drive the tiling metric.

\subsection{Field Positioning}

When 5 or more fibres are not assigned, despite there being 600 potential targets, the central coordinates of the tile are moved to a more favourable position (we find this situation occurs for $\sim 25\%$ of tiles). The most successful approach is to take the median right ascension and declination of all targets, and use this as the new tile centre. This overcomes the effects of unusual geometries (even with the region buffer, corners can be a problem), without allowing outliers to unduly influence the tile centre. If a shift in tile centre is required, then the survey buffer is no longer used (hence the small fraction of field centres inside the buffer region in Figure \ref{gamatiles}).

The final adjustment to the tiling algorithm is an option to force a tile to lie within a certain RA range of the GAMA region about to be observed. This This may be necessary at the start or end of a night when the only observable GAMA region is still at high airmass. When actually observing this meant the first field of the night had to be within the low RA $16^{m}$ of GAMA 09, and the last field of the night had to be within the high RA $16^{m}$ of GAMA 15.

\subsection{Survey Selection Function}

To some degree the greedy algorithms discussed, and the {\it dengreedy} algorithm used, will allow a selection function for the survey to be calculated; the algorithm is both simple and reproducible. However, due to continuous feedback from failed observations (typically due to bad weather) the survey will never be trivial to reproduce from start to finish. In the case of GAMA, the algorithm was implemented from a partially completed state, further complicating the calculation of a full selection function.

Ultimately, varying instrument efficiency (especially over multiple years), seeing, throughput as a function of plate position and weather will all conspire to make the true selection unknown, and any retrospective calculation an approximation. This is even assuming {\sc configure} behaves in a perfectly predictable way, but the variations in fibre assignments (especially with the addition of object feedback) will produce highly divergent tile allocations in the latter stages of the survey. As an example, when running simulations discussed above multiple runs will produce identical tile centres for the first ~20 tiles, but small deviations in coordinate positions begin to appear beyond this point. By the last few tiles of the survey the distribution of targets can differ entirely. This is indicative of the complex, and unavoidable, interplay between fibre distributions on plates and plate distributions on the sky, and clearly a perfect selection function is limited by the precise behaviour of {\sc configure}.

GAMA aims to overcome the worst aspects of an uncertain selection function by achieving unprecedented levels of completeness, as defined in multiple ways. If 100\% (or near to it) target completeness is achieved then all our survey statistics will be heavily dominated by cosmic (or sample) variance rather than our selection function.

\section{GAMA Survey Progress and Predictions}

In year 2 107 fields were observed (from a possible 154), which is slightly better than the median return at the AAT for that time of year, and from these 31,836 good quality ($Q\ge3$) redshifts were obtained. This is a lot less than in year 1, but largely due to unavoidable factors (weather effects and instrument downtime). Also, fainter magnitude limits were used for year 2 targets ($r<19.4$ in GAMA 09 and GAMA 15 for year 2 compared to $r<19.0$ for year 1), which obviously affects the average $S/N$ and lowers the redshift success rate.

\begin{figure}[t]
\begin{center}
\includegraphics[width=8cm]{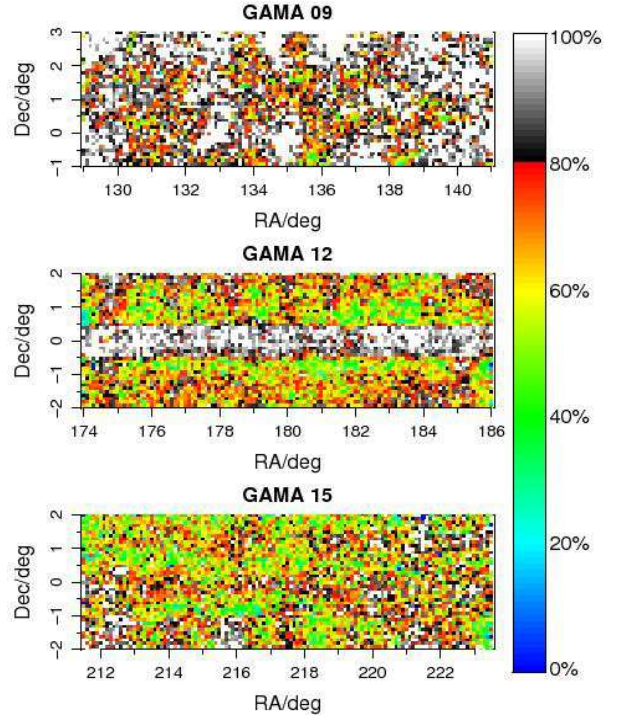}
\caption{\footnotesize The state of the GAMA regions after the second year of data. See Figure \ref{gamastart} for further details of the completeness metric.}
\label{gamaendY2}
\end{center}
\end{figure}

Due to a mixture of observational constraints and a keenness to progress one field to the point where halo mass function science is possible, GAMA 09 had 40 of these fields, GAMA 12 had 42 whilst GAMA 15 only had 23. The spatial completeness maps for each GAMA region after the completion of the second year GAMA observations are shown in Figure \ref{gamaendY2}.

It is clear from these plots that GAMA 09 is the nearest to achieving the spatial completeness target for GAMA. In fact GAMA 09 is just over 95\% complete for the main survey after year 2, and over 93\% spatially complete (using the earlier definition of what fraction of the region achieves 80\% local completeness). G12 is 83\% complete for the main survey and 66\% spatially complete. G15 is 82\% complete for the main survey and 65\% spatially complete.

From the current state of the GAMA survey for all 3 regions, it is possible to make quite accurate predictions about how the survey will appear after the third and final year of observations assuming particular weather losses. The expectation at the AAT is that there is a 2/3 probability of a given field being successfully observed. Due to the observational constraints of the survey we expect $\sim154$ fields to be observed (this was the field limit for the GAMA year 2 time allocation, due to fitting observations around dark-time the number of year 3 fields will differ). Assuming a binomial distribution for the probability of fields being observed, the median number of successful fields we expect in year 3 is 103 (slightly less than the number obtained in year 2). We define ``weather minus 1 sigma'' to be the number of tiles for which the integrated binomial distribution is equal to the integrated normal distribution from $-\infty$ to $-1\sigma$ (0.159): this equates to 97 tiles. Based on similar logic we can calculate that year 2 had $+0.5\sigma$ weather, and year 1 and year 2 combined had better than $+5\sigma$ weather (mostly due to the near perfect weather during year 1). Using these numbers for available tiles, we can make reasonable, conservative, predictions for the final state of the first 3 years of the GAMA survey.

\begin{figure}[t]
\begin{center}
\includegraphics[width=8cm]{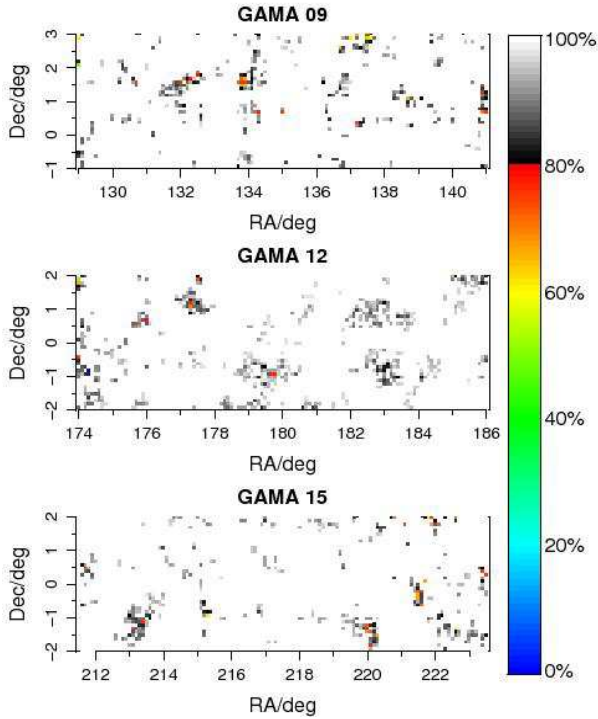}
\caption{\footnotesize The predicted state of the GAMA regions when all the major survey targets have been achieved. This requires 12 more tiles for G09, 32 more tiles for G12 and 27 more tiles for G15. See Figure \ref{gamastart} for further details of the completeness metric.}
\label{tarY3}
\end{center}
\end{figure}

To achieve the hardest survey requirement of 99\% completeness in each GAMA region will take a total of 71 more fields (in practice a few more will be required when bad weather failures are fed back in). This is well inside even the weather $-1\sigma$ limit, and requires 12 more tiles for G09, 32 more for G12 and 27 more for G15. Figure \ref{tarY3} shows what the local completeness maps for each region will look like at the exact point all our main survey targets are achieved. Encouragingly even GAMA 12 looks uniformly complete, despite the deep strip created during year 1 making the distribution of tiles (and fibres) much less random in declination.

\begin{figure}[t]
\begin{center}
\includegraphics[width=8cm]{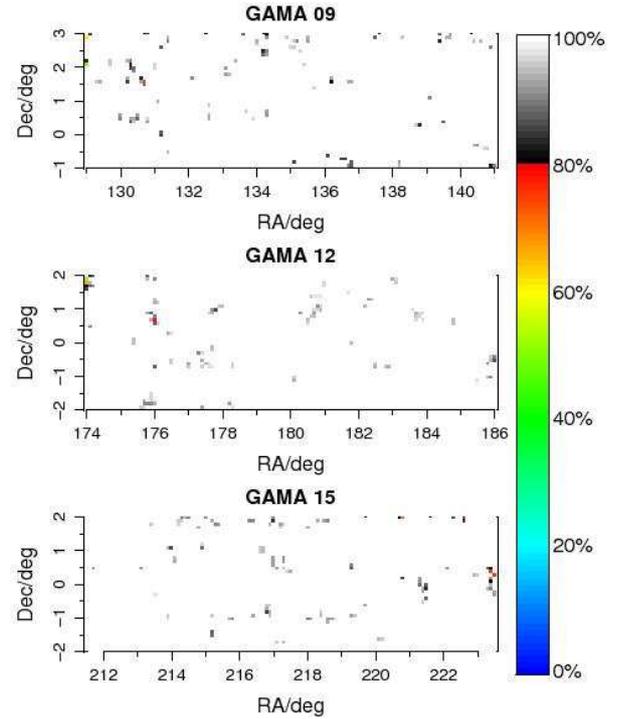}
\caption{\footnotesize The predicted state of the GAMA regions assuming we experience weather $-1\sigma$ for year 3. This provides 20 more tiles for G09, 42 more tiles for G12 and 35 more tiles for G15 (i.e.\ 97 tiles are predicted, and these are distributed according to available targets). See Figure \ref{gamastart} for further details of the completeness metric.}
\label{m1sY3}
\end{center}
\end{figure}

Figure \ref{m1sY3} shows the completeness maps for each GAMA region assuming weather $-1\sigma$ (i.e.\ 97 more tiles), thus is a fairly conservative estimate of the final state of the survey. This would provide 20 tiles for G09, 42 tiles for G12 and 35 tiles for G15 if similar completeness is desired in all three GAMA regions. It is not unrealistic for the hardest survey goal to be reached either---the requirement that every 0.1 magnitude Petrosian $r$-band bin is 99\% complete. This is predicted to take 110 more tiles, and would require a slightly better run than experienced in year 2 (107 tiles). In terms used previously, this would require weather $+1.3\sigma$.

Since bad weather is often highly correlated, it is optimistic to treat the chance of consecutive fields being observed as independent events (as required for a true binomial distribution). However, even in the extreme event of half the fields being lost due to bad weather we will still achieve our primary survey requirements, and even one week of perfect weather should be enough to bring all three GAMA regions over the 95\% completeness mark.

\begin{figure}[t]
\begin{center}
\includegraphics[width=8cm]{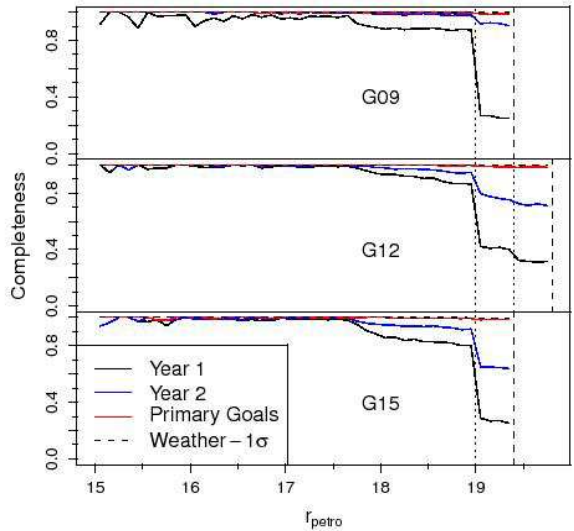}
\caption{\footnotesize Completeness as a function of $r_{petro}$ at the end of year 1, year 2 and the predicted state if primary goals are met or we experience weather $-1\sigma$. The completeness is a strong function of $r_{petro}$, and it is clear that fainter bins suffer worst. The secondary target of achieving 99\% completeness in all 0.1 mag will take 110 tiles, or weather $+1.3\sigma$. Black dotted lines indicate an interim survey limit, causing the steps seen after year 1. Black dashed lines indicate the ultimate survey limit as stated in the GAMA survey requirements. The data is limited to the main survey sample, so fainter filler targets do not contribute.}
\label{rpetcomp}
\end{center}
\end{figure}

Figure \ref{rpetcomp} shows how the completeness of each GAMA region varies as a function of $r_{petro}$, where fainter targets are less complete. Achieving a uniform level of 99\% completeness over all magnitudes is clearly a harder task than achieving 99\% global completeness, and since redshift success depends on apparent magnitude (brighter objects have a higher success rate) we should also expect the failure rate to increase from that experienced so far. This will mean extra fields will be required to obtain safe redshifts, and these numbers should be considered lower limits for survey predictions.

\section{Summary}

This work has demonstrated that a greedy approach to tiling proves to be extremely successful in densely packed surveys such as GAMA. By aggressively targeting under-densities with each field used, high levels of spatial completeness should be a reality for each GAMA region by the end of the third year of observations. In the meantime we allow for a simple mechanism to feed redshift failures back in, and by prioritising highly clustered regions we obtain both a large number of close pairs, and guarantee we are not left with difficult pockets of galaxies in the final stages of the survey. Further still, by utilising every non-main survey fibre on deeper targets, we ensure efficient use of the 2dF instrument, and make a head-start on any future extended redshift surveys in the GAMA regions. 

\section*{Acknowledgments} 
ASGR acknowledges STFC funding for the GAMA post-doctoral fellowship. Thanks also to Keith Shortridge for incorporating requested changes to {\sc configure} (creating v7.10+) that were vital to the success of the tiling algorithm used for the GAMA survey.

\end{document}